\documentclass[11pt]{article}

\usepackage{a4wide}
\usepackage{mathrsfs}
\usepackage[T1]{fontenc}
\usepackage{mathpazo}
\usepackage{setspace}
\usepackage{amsfonts}
\usepackage{amssymb}
\usepackage{amsmath}
\usepackage{epsfig}
\usepackage{latexsym}
\usepackage{color}
\usepackage{nicefrac}
\usepackage[latin1]{inputenc}

\def\crbig{\\\noalign{\vspace{1.1mm}}}

\author{
  \begin{minipage}{.97\linewidth}
    \vspace{1cm}
    \begin{center}
      \begin{small}
        \textbf{P.M. Petropoulos}\footnote{marios@cpht.polytechnique.fr}
      \end{small}
    \end{center}
    \vspace{0.5cm}
    \hspace{2cm}\begin{minipage}{.7\linewidth}
     {\it \begin{footnotesize}
     \begin{center}
         Centre de Physique Th\'eorique, CNRS\footnote{Unit\'e mixte UMR7644.},
        Ecole Polytechnique, \\
        91128 Palaiseau Cedex, France
        \end{center}
     \end{footnotesize}}
    \end{minipage}
    \vspace{0.5cm}
  \end{minipage}
}

\date{December 2007}

\title{\vspace{0.75cm}
 \boldmath \begin{huge}
    \textbf{Non-unimodular reductions and $N=4$ gauged supergravities}
  \end{huge} \unboldmath
}

\begin{document}

\begin{titlepage}
  \maketitle
  \thispagestyle{empty}

  \vspace{-9.5cm}
  \begin{flushright}
    CPHT-PC158.1107
  \end{flushright}

  \vspace{10cm}

  \begin{center}
    \textsc{Abstract}\\
  \end{center}
We analyze the class of four-dimensional $N=4$ supergravities
obtained by gauging the axionic shift and axionic rescaling
symmetries. These theories are formulated with the machinery of
embedding tensors and shown to be deducible from higher dimensions
using a Scherk--Schwarz reduction with a twist by a non-compact
symmetry. This allows to evade the usual unimodularity requirement
and completes the dictionary between heterotic gaugings and
fluxes, at least for the ``geometric sector''.

\vspace{3cm} \noindent Based on works with J.-P. Derendinger and
N. Prezas. To appear in the proceedings of the RTN workshop
\textsl{Constituents, fundamental forces and symmetries of the
universe}, Valencia, Spain, October 1 -- 5 2007.

\end{titlepage}

\onehalfspace

\tableofcontents

\newpage

\section{Why gaugings and fluxes?}\label{sec:motsum}

String compactifications share a set of usual caveats. First
stands the issue of supersymmetry breaking. The original $N=4$ or
$8$ supersymmetry of type II, heterotic or M theory vacua must be
reduced to $N=1$ at some reasonably low scale before being broken
completely to match with TeV-scale phenomenology. A second major
problem is the issue of moduli stabilization: many massless
neutral scalars pollute the spectrum and disable any attempt of
confrontation with low-energy physics. The third problem is that
of the cosmological constant. It is obviously related to the
previous ones, the common denominator of all these being the
structure of the vacuum. It is not clear though to what extent
string theory can shed light on this infrared problem.

The possibility to give vacuum expectation values to
antisymmetric-tensor fields (NS--NS, R--R, spin connection)
provides a tool for a better control of the situation. This was
recognized long ago and has been reexamined extensively over the
recent years (see \cite{Grana:2005jc} for a comprehensive review)
with essentially two complementary approaches. The connection
between these two methods (see \cite{Trigiante:2007ki} for a
concise review) stems from the fact that the effective theories of
flux compactifications are \emph{gauged} supergravities with
spontaneously broken supersymmetry, with scalars charged under
(non-)Abelian gauge groups, and with moduli-dependent
superpotential (and potential) -- to be opposed to toroidal
compactifications, which are \emph{ungauged} supergravities with
neutral scalars and flat potential. The first approach might be
called ``top-down'' and consists in (\romannumeral1) understanding
the generalized geometrical tools that describe the
ten-dimensional theory in presence of fluxes, (\romannumeral2)
find admissible compactifications, and (\romannumeral3) analyze
the low-energy properties -- in other words check the issues of
stabilization and supersymmetry breaking, not a priori guaranteed.
The second, somehow less popular (see
\cite{Derendinger:2004jn,Derendinger:2005ph} and
\cite{Derendinger:2006jb} for a review), has four dimensions as
starting point. In this ``bottom-up'' scheme (\romannumeral1) one
starts directly with phenomenologically relevant four-dimensional
gauged supergravities, (\romannumeral2) one translates the gauging
parameters into fluxes, and (\romannumeral3) one tries to
reconstruct the fundamental theory. The latter point might be
subtle because no systematic oxidation recipe exists and not all
four-dimensional gauged $N=4$ ($N=8$) supergravities are
heterotic, type-I or type-II-orientifold (M-theory) vacua. Despite
this reservation, the method has been shown to capture a large
variety of situations, including four-dimensional remnants of
non-geometric string backgrounds or supersymmetric
$\mathrm{AdS}_4$ vacua of type II theories with stabilized main
moduli \cite{Derendinger:2004jn,Kounnas:2007dd}.

Here we will focus on four-dimensional $N=4$ theories
and remind the basics on the gauging
procedure using the \emph{embedding tensor} -- outstanding tool
described e.g. in \cite{deWit:2002vt,
deWit:2003hq,deWit:2003hr,deWit:2004nw,deWit:2005ub,Schon:2006kz}.
We will in particular analyze the gauging of \emph{axionic shifts
and rescalings} and trace its ten-dimensional origin. This is not
straightforward: it requires a generalized, \emph{non-unimodular}
Scherk--Schwarz reduction with a twist by the scaling symmetries
and relies on a duality between massive vectors and massive
two-forms.

\section{Gauged supergravities and the embedding tensor}\label{sec:gausugr}

The ungauged four-dimensional $N=4$ supergravity has in general
$1$ gravitational and $n$ vector multiplets. The bosonic content
of the gravitational multiplet is $1$ graviton, $6$ graviphotons,
and $2$ real scalars combined into the \emph{axion-dilaton} $\tau
= \chi + i \exp -2\phi$; the vector multiplet has $1$ vector and
$6$ real scalars. The gauge group is Abelian, $U(1)^{6+n}$, and
all scalars are neutral and non-minimally coupled to the vectors
(interaction terms of the type $f(\mathrm{scalars})\, F^2$). There
is \emph{no} scalar potential.

The elimination of the auxiliary fields generates the \emph{scalar
manifold:}
\begin{equation}\label{eq:scalman}
  \mathcal{M}=\frac{SL(2,\mathbb{R})}{U(1)}\times \frac{SO(6,n)}{SO(6)\times
  SO(n)},
\end{equation}
which exhibits the global symmetries of the theory. The
$SL(2,\mathbb{R}) \times  SO(6,n)\subset Sp(12+2n, \mathbb{R})$ is
realized as a U-duality symmetry of the \emph{full} theory. In
heterotic theory, only an $A_{2,2} \times  SO(6,n)$ is realized
off-shell. The $A_{2,2}$ generates the axionic rescaling and
axionic shifts and does not mix electric and magnetic gauge
fields. Genuine electric--magnetic duality transformations relate
different Lagrangians written in different ``symplectic frames''.

Although one set of vectors only describes propagating degrees of
freedom -- electric or magnetic or any combination depending on
the choice of symplectic frame --, it is possible to include them
all in a unified Lagrangian formulation. The latter comprises
$12+2n$ fields $\left(\left\{\mathbf{A}^{M+}\right\},
\left\{\mathbf{A}^{M-}\right\}\right), M=1,\ldots,6+n$, which form
a $(\mathbf{2}, \mathbf{Vec})$ of $SL(2,\mathbb{R})\times SO(6,n)$
(i.e. a $\mathbf{Vec}$ of $Sp(12+2n, \mathbb{R})$), without
kinetic term for $\left\{\mathbf{A}^{M-}\right\}$. It also
includes extra two-form auxiliary fields dual to the scalars. The
equations of motion for the magnetic vectors and the two-forms set
the duality between the scalars and the two-forms, and between the
electric and magnetic vectors, respectively. The presence of all
auxiliary fields (magnetic vectors \emph{and} two-forms) is also
necessary for gauge invariance, which ensures the decoupling of
all ghosts. The number of propagating degrees of freedom remains
unaltered, as compared to the ungauged theory.

The gauging of a supergravity theory is a deformation which is
compatible with supersymmetry. For $N=4$ (and $N=8$ which we will
not discuss here) this is the only possible deformation. It is a
promotion of a subgroup of the U-duality group to a local gauge
symmetry supported by (part of) the \emph{existing} $U(1)^{n+6}$
vectors.

The generators of the duality group are
    $T^{MN}=-T^{NM},\ M,\ldots = 1, \ldots, 6+n$ for
the $SO(6,n)$ subgroup and
   $S^{\beta\gamma}= S^{\gamma\beta},\ \beta, \ldots = +,- $ for
the $SL(2,\mathbb{R})$. They obey the following commutation
relations:
\begin{equation}
\label{eq:grcom}
\begin{array}{rcl}
\displaystyle{\left[T_{KL},T_{JM}\right]}&=&\displaystyle{\eta_{LJ}
T_{KM}+\eta_{KM} T_{LJ}-\eta_{KJ} T_{LM} - \eta_{LM} T_{KJ}}
\crbig
\displaystyle{\left[S_{\alpha\beta},S_{\gamma\delta}\right]}&=&\displaystyle{-
  \epsilon_{\alpha\gamma}\, S_{\beta\delta} -
  \epsilon_{\beta\delta}\, S_{\alpha\gamma} -
  \epsilon_{\alpha\delta}\, S_{\beta\gamma} -
  \epsilon_{\beta\gamma}\, S_{\alpha\delta}}
\end{array}
\end{equation}
%
with $\eta_{LJ}$ being the $SO(6,n)$-invariant metric, and
$\epsilon^{+-}=1=\epsilon_{+-}$\footnote{Indices $M,N, \ldots$ are
lowered and raised with $\eta_{LJ}$ and $\eta^{KM}$ (inverse
matrix). With the present conventions for
$\epsilon_{\alpha\beta}$, $\epsilon_{\gamma\alpha}\,
\epsilon^{\gamma\beta} = \delta_\alpha^\beta$, we can raise and
lower $\alpha$-indices unambiguously as follows: $A_\alpha=
A^\beta \, \epsilon_{\beta\alpha}$ and $B^\alpha =
\epsilon^{\alpha\beta}B_\beta$. This leads to $A_+=-A^-$ and
$A_-=A^+$. In particular, $S_{++} = S^{--}$, $S_{+-} = -S^{+-}$
and $S_{--} = S^{++}$. }. The generators of the \emph{gauge
algebra} are
\begin{equation}\label{eq:embed}
  \Xi_{\alpha L} = \frac{1}{2}\left(\Theta_{\alpha LMN}\, T^{MN} +
  \Theta_{\alpha L\beta\gamma}\, S^{\beta\gamma}\right),
\end{equation}
where $\{\Theta_{\alpha LMN}, \Theta_{\alpha L\beta\gamma} \} \in
(\mathbf{2}, \mathbf{Vec} \times \mathbf{Adj}) + (\mathbf{2}\times
\mathbf{3}, \mathbf{Vec} )$ of $SL(2,\mathbb{R})\times SO(6,n)$ is
the \emph{embedding tensor}. Both electric and magnetic gauge
transformations are included in this formalism. However, as we
already emphasized, at most $6+n$ $\Xi$'s are independent. Put
differently, the embedding tensor has maximal rank $6+n$ and is
therefore subject to \emph{constraints} that we will now briefly
discuss -- for a comprehensive exposition, we recommend the
already quoted literature.

Demanding supersymmetry leads to a set of \emph{linear
constraints}. This reduces the embedding tensor to $(\mathbf{2},
  \mathbf{Ant}_{[3]}) + (\mathbf{2}, \mathbf{Vec} )$:
\begin{equation}\label{eq:redemb}
\Xi_{\alpha L} = \frac{1}{2} \left( f_{\alpha LMN}\, T^{MN}
+\eta_{LQ}\,\xi_{\alpha P} \,
  T^{QP}+ \epsilon^{\gamma\beta}\,\xi_{\beta L}\,
  S_{\gamma\alpha}\right),
\end{equation}
where $f_{\alpha LMN}\equiv f_{\alpha [LMN]}$ and $\xi_{\beta L}$
are the irreducible blocks that define the embedding tensor. They
allow for the complete determination of the gauge algebra and its
commutators, the charges and covariant derivatives, the scalar
potential and the mass matrices.

Since the vectors belong to the fundamental\footnote{This means in
particular that the components $\Xi_{\alpha L\, \beta
M}^{\hphantom{\alpha L\, \beta M}\gamma N}$ of $\Xi_{\alpha L}$,
as they can be read off from Eq. (\ref{eq:redemb}), are not
entries of an antisymmetric $(12+2n)\times (12+2n)$ matrix.
Consistency in gauge transformations is nevertheless ensured by
the introduction of the auxiliary two-forms.} of $Sp(12+2n,
\mathbb{R})$, we must also impose that the latter contains the
adjoint of the gauge algebra, and that this algebra closes. The
minimal set of \emph{quadratic} constraints that allow to fulfill
these requirements is the following:
\begin{equation}
\label{eq:firstfourthcon}
  \begin{cases}
   \displaystyle{\eta^{MN}\,\xi_{\alpha M}\, \xi_{\beta N}=0} & \mathrm{(i)}
  \crbig
   \displaystyle{\eta^{MN}\,\xi_{(\alpha M}\, f_{\beta) NIJ}=0} & \mathrm{(ii)}
  \crbig
   \displaystyle{\epsilon^{\alpha\beta}\left(\xi_{\alpha I }\, \xi_{\beta J} +
  \eta^{MN}\,\xi_{\alpha M} \,f_{\beta N I J} \right)=0} & \mathrm{(iii)}
  \crbig
  \displaystyle{\eta^{MN}\, f_{\alpha MI[J}\, f_{\beta KL]N}
  \frac{1}{2}\xi_{\alpha [J}\,f_{\beta KL]I} -} &
  \crbig
  \displaystyle{- \frac{1}{6}\epsilon_{\alpha\beta}\, \epsilon^{\gamma\delta}\, \xi_{\gamma I}\, f_{\delta JKL}+
  \frac{1}{2}\eta^{MN}\, \xi_{\alpha M}\, f_{\beta N[JK}\,\eta_{L]I} + \frac{1}{6}f_{\alpha JKL}\,
  \xi_{\beta I}=0.} & \mathrm{(iv)}
  \end{cases}
\end{equation}
The \romannumeral4$^{\mathrm{th}}$ constraint is Jacobi-like. It
should be stressed, however, that $f_{\alpha JKL}$ \emph{are not
necessarily} structure constants of some algebra. The structure
constants of the gauge algebra are given in general in terms of
all gauging parameters i.e. $f_{\alpha JKL}$ and $\xi_{\beta L}$.

Most of the known solutions to the above set of constraints have
vanishing $\xi_{\alpha L}$. In this case only the
\romannumeral4$^{\mathrm{th}}$ constraint in Eqs.
(\ref{eq:firstfourthcon}) survives. This ``generalized Jacobi''
identity can be solved by introducing \emph{de Roo--Wagemans
phases} \cite{deRoo:1985jh} that relate the electric ($f_{+ LMN}$)
and magnetic ($f_{- LMN}$) gauging parameters \cite{Schon:2006kz}.
Hence, the solutions correspond to pure $SO(6,n)$ gaugings,
extensively studied in the literature, which lead to a variety of
gauge algebras such as \emph{flat algebras}. From a
higher-dimensional perspective the latter turn out to be related
to \emph{unimodular} Scherk--Schwarz reductions\footnote{Flat
groups were introduced in the reduction scheme proposed in
\cite{Scherk:1979zr}, as the only solution to a double
requirement: (\romannumeral1) scalar potential bounded from below
and (\romannumeral2) vanishing of the cosmological constant.} (see
e.g. \cite{Andrianopoli:2002mf}), in contrast to what we will be
discussing in the following.

Our aim is here to elaborate on gaugings with non-vanishing
parameters $\xi_{\alpha L}$. Only a few isolated examples have
been studied so far that fall in this class
\cite{Villadoro:2004ci,Schon:2006kz} and their more systematic
analysis will be the subject of the next chapters.

\section{The axionic transformations and their gaugings}\label{sec:nonunim}

The axionic transformations are generated by a subgroup of the
$SL(2,\mathbb{R})$. The latter acts on the axion-dilaton as a
Möbius transformation: $\tau \to  \nicefrac{a\tau+b}{c\tau+d}$.
Since the axion-dilaton parameterizes the $SL(2,\mathbb{R})/U(1)$
coset, we can define a $2\times 2$ matrix,
\begin{equation}\label{eq:linaxdil}
  \left(M^{\alpha \beta}\right) =\frac{1}{\textsf{Im} \tau}
    \begin{pmatrix}
    1& -\textsf{Re} \tau\\
    -\textsf{Re} \tau&|\tau|^2
    \end{pmatrix},
\end{equation}
on which $SL(2,\mathbb{R})$ acts linearly, with matrices $\begin{pmatrix} a&b\\
c&d\end{pmatrix}$. The generator of the genuine electric--magnetic
duality is $S^{--}=\begin{pmatrix} 0&0\\
2&0\end{pmatrix}$, whereas axionic shifts  $\tau \to \tau +b $ and
axionic rescalings $\tau \to a^2 \tau$ are generated by $S^{++}=\begin{pmatrix} 0&-2\\
0&0\end{pmatrix}$ and $S^{+-}=\begin{pmatrix} 1&0\\
0&-1\end{pmatrix}$, respectively.

It is clear form Eq. (\ref{eq:redemb}), that gauging the axionic
symmetries requires an embedding tensor with $\xi_{\alpha M}\neq
0$. Such a gauging is necessarily accompanied by a partial gauging
of  $SO(6,n)$. As already advertised, our motivation is to proceed
with \emph{electric gaugings}, i.e. gaugings of the axionic shifts
$S^{++}$ and rescalings $S^{+-}$ but \emph{not} the
electric--magnetic duality transformation $S^{--}$. We must
therefore set $\xi_{-I}=0$. Our further choice is $f_{-LMN}=0$.
Although this is not compulsory for general electric gaugings, it
simplifies considerably the quadratic constraints for $\xi_{+I},
f_{+LMN}$ (the ``$+$'' index will be dropped from now on). For
simplicity, we focus on the case $n=6$ corresponding to 12 vectors
in total, and use the light-cone-like convention:
$\{I\}\equiv\{i,i'\}$,
\begin{equation}\label{eq:lceta}
\eta =  \begin{pmatrix}
   0 & \mathbb{I}_6\\
   \mathbb{I}_6 & 0
    \end{pmatrix}.
\end{equation}

A non-trivial solution to Eqs. (\ref{eq:firstfourthcon}) is
captured by six real parameters, $\{\lambda_{i}, i=1,\ldots,6\}$:
\begin{equation}\label{eq:nonunisol}
\xi_i= \lambda_i, \quad \xi_{i'}=0, \quad
f_{i'ij}=f_{ji'i}=f_{iji'}=- \lambda_{[i}\, \delta_{j]i'},\quad
f_{ijk}=f_{ii'j'}=f_{i'j'k'}=0.
\end{equation}
Several remarks are in order here. The gauging under consideration
will be called ``non-unimodular'' for reasons that will become
clear later, or ``tracefull'' since
\begin{equation}\label{eq:trace}
  f_{ij}^{\hphantom{ij}j}=-\frac{5}{2}\lambda_{i}^{\hphantom{j}}.
\end{equation}
This is slightly misleading because the gauge algebra \emph{is}
traceless as a consequence of the full antisymmetry of its genuine
structure constants. The latter \emph{are not}
$f_{ij}^{\hphantom{ij}k}$, which are not Lie-algebra structure
constants, but specific combinations of $f_{IJK}$ and $\xi_I$,
read off from the commutation relations of generators
(\ref{eq:redemb}). We find in this way 8 independent generators
out of $2\times 12$, as we see from the following:
\begin{equation}
\label{eq:8gene}
\begin{array}{rcl}
 &&\displaystyle{\Xi_{-i}= - \frac{\lambda_{i}}{2}S^{++}\equiv
 \lambda_{i} \Xi}  \crbig
 &&\displaystyle{\Xi_{-i'}=0} \crbig
 &&\displaystyle{\Xi_{+i}=- \frac{\lambda_{i}}{2}
 \left(T_{\hphantom{j}j}^{j} +    S^{+-} \right)\equiv \lambda_i\Upsilon}  \crbig
&&\displaystyle{\Xi_{+i'}^{\vphantom{j}}=-\lambda_j^{\vphantom{j}}
        T_{\hphantom{j}i'}^{j}\equiv \Xi_{i'}^{\vphantom{j}}}
\end{array}
\end{equation}
%
The commutation relations for $\{\Upsilon,\Xi,\Xi_{i'}\}\subset
SL(2,\mathbb{R})\times SO(6,6)$ are
\begin{equation}
\label{eq:8commu}
\begin{array}{rcl}
 \displaystyle{ \left[\Xi_{i'}, \Xi_{j'}\right]}&=&\displaystyle{0} \crbig
 \displaystyle{ \left[\Xi, \Xi_{j'}\right]}&=&\displaystyle{0} \crbig
 \displaystyle{\left[\Xi_{i'}, \Upsilon \right]}&=&\displaystyle{\Xi_{i'}}\crbig
 \displaystyle{ \left[\Xi,\Upsilon\right]}&=&\displaystyle{- \Xi} .
\end{array}
\end{equation}
%
In contrast to the algebras obtained by standard Scherk--Schwarz
reductions, the one under consideration is \emph{non-flat}. The
$\{\Upsilon, \Xi \}$ is the non-semi-simple subalgebra
$A_{2,2}\subset SL(2,\mathbb{R})$ of axionic rescalings and
axionic shifts. These axionic symmetries are gauged along with $6$
Abelian generators $\{\Xi_{i'}\}\subset SO(6,6)$, and
$\{\Upsilon,\Xi_{i'}\}$ spans a seven-dimensional ideal. Notice
finally that one could possibly introduce non-abelianity by
switching on $f_{ijk}$ provided $f_{[ijk}\lambda_{\ell]}=0$, as
follows from  (\ref{eq:firstfourthcon}).

\section{Dynamics of the axionic gaugings}\label{sec:dynamics}

We would like now to discuss some dynamical aspects of the axionic
gaugings. For this we need to describe the Lagrangian formulation
of the system -- including electric and magnetic components.
Following \cite{Schon:2006kz}, the bosonic sector of the
Lagrangian associated with any consistent gauging, given in terms
of the parameters $f_{\alpha JKL}$ and $\xi_{\beta L}$, has three
parts:
\begin{description}
\item[-] $\mathcal{L}_{\textsf{kin}}$: kinetic terms for graviton, electric vectors and
scalars,
\item[-] $\mathcal{L}_{\textsf{top}}$: auxiliary-field contributions (magnetic vectors and two-forms)
necessary to maintain the correct number of propagating fields,
\item[-] the scalar potential:
\begin{eqnarray}
 {\cal L}_{\textsf{pot}} &=&-\frac{e}{16} \bigg\{f_{\alpha
    MNP}\, f_{\beta QRS}\, M^{\alpha\beta}\left(\frac{1}{3} M^{MQ}\,
    M^{NR}\, M^{PS} + \left(\frac{2}{3} \eta^{MQ} -
    M^{MQ}\right)\eta^{NR} \eta^{PS}\right)\nonumber \\
    &&-\frac{4}{9} f_{\alpha MNP} \,
    f_{\beta QRS}\, \epsilon^{\alpha\beta} \, M^{MNPQRS} + 3
    \xi_{\alpha}^M\, \xi_{\beta}^N\, M^{\alpha \beta} M_{MN}
    \bigg\},
  \label{eq:genpot}
\end{eqnarray}
where $e$ is the vierbein,  $M^{\alpha\beta}$ is given in
(\ref{eq:linaxdil}) and $M^{MQ}$ parameterizes similarly the
remaining $36=21+15$ scalars of $\nicefrac{SO(6,6)}{SO(6)\times
SO(6)}$:
\begin{equation}\label{eq:lin36scal}
  M^{MN}=\begin{pmatrix}
    h^{ij} &  - h^{ik}\, b_{kj} \\
    b_{ik}\, h^{kj}  &  h_{ij}-b_{ik}\, h^{k\ell} \, b_{\ell j}
  \end{pmatrix};
\end{equation}
$M^{MNPQRS}$ is another tensor build out of the $36$ scalars (see
\cite{Schon:2006kz} or \cite{Derendinger:2007xp} for precise
expressions).
\end{description}

Working out the kinetic terms for the axionic gauging, it appears
that among the original $12$ vectors, $4$ remain inert while $2+6$
are embedded in $SL(2,\mathbb{R})\times SO(6,6)$ as generators of
local symmetries -- they enter in covariant derivatives acting on
scalars. One can also obtain the specific scalar potential by
inserting (\ref{eq:nonunisol}) into the general expression
(\ref{eq:genpot}):
\begin{equation}\label{eq:axpot}
  {\cal L}_{\textsf{pot}}= \frac{1}{16}  \textsf{e}^{2 \phi} \lambda_i \left( 8 h^{ij}-
    h^{ij}\, h^{k\ell}\, b_{\ell m}\, h^{mn}\, b_{nk} + 2 h^{ik}\, b_{km}\, h^{mn}\,
    b_{nr}\, h^{rj}\right) \lambda_j.
\end{equation}
This is positive definite (as was the scalar potential in
\cite{Villadoro:2004ci}).

The dynamics of the axion-dilaton requires a careful treatment.
The kinetic term for this field is
\begin{equation}\label{eq:kinaxdil}
e^{-1} {\cal L}_{\textsf{kin:axion-dilaton}}=- D_\mu \, \phi D^\mu
  \phi - \frac{1}{4} \textsf{e}^{4 \phi} D_\mu \chi \, D^\mu
  \chi,
\end{equation}
where
\begin{equation}\label{eq:covdevaxdil}
D_\mu \phi = \partial_\mu \phi -\frac{1}{2} Y_\mu \quad
\mathrm{and} \quad D_\mu \chi =\partial_\mu \chi+  X_\mu + Y_\mu
\chi
\end{equation}
are the covariant derivatives involving the physical vectors
\begin{equation}\label{ea:physvec}
Y_\mu^{\vphantom{+}} = \lambda_i^{\vphantom{+}} \,
  A_\mu^{i+}\quad \mathrm{and} \quad  X_\mu^{\vphantom{i}} = \lambda_i^{\vphantom{i}} \,
  A_\mu^{i-}.
\end{equation}
It is worth stressing that these vectors are combinations of
\emph{both} electric \emph{and} magnetic potentials. The gauging
mixes the spurious and physical fields, keeping the number of
propagating degrees of freedom fixed, though. The vectors $Y_\mu$
and $X_\mu$ are associated with the gauging of the symmetries
generated by $\Upsilon$ (axion rescalings, $\chi \to a^{2} \chi$,
$\phi\to \phi - \log a$) and $\Xi$ (axion shifts, $\chi \to \chi +
b$) respectively. Because of these local symmetries, the axion can
be gauged away. In this process, $X_\mu$ acquires a mass via its
Stückelberg coupling to $\chi$ and can be traded for a massive
two-form $C_{\nu\rho}$\footnote{Trading massive vectors for
massive two-forms in four dimensions has been used in various
instances \cite{Townsend:1983xs,Quevedo:1995ep}.}.

The final bosonic content of the axionic gauging is as follows:
the dilaton, $4+1+6$ vectors with Abelian algebra ($4$ inert, $1$
associated with the axionic rescalings of $SL(2,\mathbb{R})$, $6$
associated with maximal-Abelian-subalgebra transformations of
$SO(6,6)$ -- translations), $1$ massive two-form and $36$ scalars
minimally coupled to the $1+6$ vectors, with scalar potential
(\ref{eq:axpot}). In general, depending on the specific values of
the parameters $\lambda_i$, more scalars can be gauged away while
vectors can simultaneously become massive thanks to their
Stückelberg couplings. Some of the remaining scalars are massive,
while other are massless.

\section{The higher-dimensional origin: non-unimodular Scherk--Schwarz reduction}\label{sec:genschsch}

We will now perform a generalized dimensional reduction of
heterotic ten-dimensional pure supergravity and show that the
resulting effective theory belongs to the class of $N=4$
four-dimensional gauged supergravities studied in Secs.
\ref{sec:nonunim} and \ref{sec:dynamics} .

The action of the heterotic ten-dimensional pure supergravity
(bosonic sector) reads:
\begin{equation}\label{eq:act10d}
 S = \int_{M_4} \mathrm{d}^4x \int_{K_6} \mathrm{d}^6y \, \sqrt{-G}\,  \mathrm{e}^{-\Phi}
     \bigg(R+G^{MN}\, \partial_M \Phi\,
     \partial_N \Phi-\frac{1}{12} H_{MNK} H^{MNK}
     \bigg),
\end{equation}
where $\Phi$ is the dilaton, $G_{MN}$ the metric and
$H=\mathrm{d}B$ the NS-NS field strength, all in  ten dimensions
and in the sigma-model frame. Since $K_6$ is compact, from the
four-dimensional viewpoint this action describes the dynamics of
an infinitude of modes. A consistent reduction provides an
effective theory on $M_4$ for a \emph{finite} subset of modes.
This requires a ``good'' choice for $K_6$ plus an ansatz for the
$y$-dependance of all fields, which sets the mode-selection
pattern. A necessary consistency condition is that $\mathcal{L}$
be $y$-independent.

For the standard reduction on a flat torus $T^6$, the ansatz is
the absence of any $y$-dependence in the ten-dimensional fields.
In this case, the bosonic spectrum consists of $1$ graviton,
$12=6+6$ Abelian vectors, $36=21+15$ scalars, $1$ dilaton and $1$
axion (dual to the NS--NS form), all massless and neutral.

The Scherk--Schwarz reduction \cite{Scherk:1979zr} is an
alternative that allows  to generate (\romannumeral1) non-Abelian
gauge symmetries and corresponding charges, (\romannumeral2) a
scalar potential and (\romannumeral3) a spontaneous breaking of
(super)symmetries.

In ordinary Scherk--Schwarz reductions, the ansatz includes a
precise $y$-dependance: around an internal cycle, the fields
transform in a way that can be reabsorbed by the action of an
internal symmetry. This is equivalent to the introduction of
geometric (spin connection) fluxes $\gamma^i_{\hphantom{i}jk}$
appearing in the exterior differentials of the vielbeins,
\begin{equation}\label{eq:spincon}
\mathrm{d}\theta^i_{\vphantom{jk}}=
         - \gamma^i_{\hphantom{i}jk}\,
         \theta^j_{\vphantom{ik}}\wedge\theta^k_{\vphantom{ji}}.
\end{equation}
These fluxes satisfy the Bianchi identity,
\begin{equation}\label{eq:bianjac}
\gamma^i_{\hphantom{i}j[k}\, \gamma^j_{\hphantom{j}\ell m]}=0,
\end{equation}
and can alternatively be considered as structure constants
\begin{equation}\label{eq:strucon}
f^{\hphantom{jk}i}_{jk}=2\gamma^i_{\hphantom{i}jk}
\end{equation}
of a group, the compact space being locally a group manifold. The
identity (\ref{eq:bianjac}) is therefore also a Jacobi identity.
The \emph{unimodularity} property,
\begin{equation}\label{eq:uni}
\gamma^i_{\hphantom{i}ij}=0
\end{equation}
is required for the consistency of the truncation (see e.g.
\cite{Scherk:1979zr,Hull:2005hk}). Many known examples fall in
this class, which include non-semi-simple or semi-simple gauge
groups -- as the \emph{twisted tori} leading to gaugings in
$SO(6,6)$.

External Scherk--Schwarz reductions are more exotic. The symmetry
which is used to guarantee the consistency of the $y$-dependence
ansatz of the fields is external. A specific example is provided
by the action (\ref{eq:act10d}), which is invariant under the
following $SO(1,1)$ scaling symmetry:
\begin{equation}\label{eq:scalsym}
  \Phi \to \Phi + 4 \lambda, \quad G_{MN} \to
    \mathrm{e}^{\lambda} G_{MN}, \quad B_{MN} \to
    \mathrm{e}^{\lambda} B_{MN}.
\end{equation}
Using this ``duality'' symmetry, one can show that the following
ansatz is consistent \footnote{The reader is referred to
\cite{Dabholkar:2002sy} for a general discussion on reductions
with duality twists.}:
\begin{equation}\label{eq:dualansatz}
  \Phi(x,y)=\Phi(x)+ 4 \lambda_i y^i, \quad G_{MN}(x,y)=\mathrm{e}^{\lambda_i y^i} G_{MN}(x), \quad
B_{MN}(x,y)=\mathrm{e}^{\lambda_i y^i} B_{MN}(x).
\end{equation}

The dynamics of this external Scherk--Schwarz reduction can be
performed in detail using the following decomposition:
\begin{equation}\label{eq:decomp}
G_{MN}\to g_{\mu\nu}, A_{\mu k}, h_{ij},\quad B_{MN}\to
B_{\mu\nu}, B_{\mu k}, b_{ij},\quad \phi=\Phi-\frac{1}{2}\log \det
\mathbf{h}.
\end{equation}
Inserting (\ref{eq:dualansatz}) and (\ref{eq:decomp}) in the
action (\ref{eq:act10d}), the $y$-dependence drops consistently
and various new features emerge: (\romannumeral1) the vectors
$A_{\mu k}$ and $B_{\mu k}$ carry Abelian gauge symmetry,
(\romannumeral2) the scalars $h_{ij}$ are charged under $A_{\mu
k}$ with charges $\lambda_k$, (\romannumeral3) the scalars
$b_{ij}$ are charged under $B_{\mu k}$ and
\emph{St\"uckelberg-coupled} to $A_{\mu k}$, (\romannumeral4) the
four-dimensional dilaton $\phi$ is \emph{St\"uckelberg-coupled} to
$A_{\mu k}$ with charges $\lambda_k$, (\romannumeral5) a scalar
potential appears for $h_{ij}$ and $b_{ij}$.

After field redefinitions and integrations one vector drops and
the two-form $B_{\mu\nu}$ becomes massive, as a consequence of the
St\"uckelberg couplings. This is indicative of the gauging of a
shift symmetry, and it is legitimate to compare the present
``duality-twisted tori'' reduction with the axionic gauging of
four-dimensional $N=4$ supergravity studied in last section. The
matching of the Lagrangians is exact and the reduced theory at
hand \emph{is} precisely the gauged supergravity of Secs.
\ref{sec:nonunim} and \ref{sec:dynamics}.

\section{Summary and outlook}\label{sec:Out}

In the framework of heterotic theory, the specific choice of
generalized Scherk--Schwarz reduction, based on the
ten-dimensional $SO(1,1)$ shift symmetry, Eq.
(\ref{eq:dualansatz}), allows to turn on the four-dimensional
gauging parameters $\xi_i$ as ten-dimensional $SO(1,1)$ shift
parameters $\lambda_i$ along the torus one-cycles. This flux
compactification is therefore equivalent to the gauging of the
four-dimensional $SL(2,\mathbb{R})$ axionic shifts and rescalings,
and makes it possible to evade the unimodularity property of more
conventional geometric fluxes: as a consequence of
(\ref{eq:trace}) and (\ref{eq:strucon}),
\begin{equation}
\label{eq:nonuni}
  \gamma_{\hphantom{j}ij}^{j}=-\frac{5}{4}\lambda_{i}^{\hphantom{j}}.
\end{equation}
As advertised, axionic gaugings are equivalent to non-unimodular
geometric fluxes. It should be stressed that the distinction
between unimodular (Eq. (\ref{eq:uni})) and non-unimodular (Eq.
(\ref{eq:nonuni})) reductions is manifest in the sigma-model
frame; in the Einstein frame all consistent reductions are
unimodular. Nevertheless, this analysis elegantly demonstrates the
power of the gauging procedure for describing diverse flux
compactifications, and closes the chapter of characterizing a
whole class of heterotic gaugings in terms of NS--NS and
spin-connection fluxes. Further oxidation to M theory is also
possible along the lines of
\cite{Dall'Agata:2005fm,Hull:2006tp,Hull:2007zu}.

Besides the precise relations that one can establish among
ten-dimensional fluxes and four-dimensional gaugings, a
fundamental and not yet unravelled question is the following: what
are the geometrical features of the fundamental theory on the top
that translate into the consistency constraints imposed to the
embedding tensor from the bottom (Eqs. (\ref{eq:firstfourthcon}))?

This question may not admit any answer, even in the framework of
``generalized geometries''. Indeed, the analysis presented in
\cite{Maharana:1992my,Kaloper:1999yr,
Derendinger:2004jn,Derendinger:2005ph,Derendinger:2007xp} calls
for further investigation of other classes of gaugings, related to
the previous by duality transformations, and corresponding to
possibly new fluxes. Following Sec. \ref{sec:gausugr}, the gauging
parameters are $f_{\alpha IJK},\xi_{\alpha L}$. These are 464 real
parameters, subject to the set of constraints
 (\ref{eq:firstfourthcon}). We can classify
them according to their nature, using the light-cone-convention
(\ref{eq:lceta}) and the subsequent splitting of the indices
$\{I\}\equiv\{i,i'\}$:
 \begin{itemize}
    \item $f_{+ IJK},\xi_{+ L}$: 232 electric parameters that
    include
       \begin{description}
         \item[-] the $f_{+ ijk}$'s corresponding to NS--NS fluxes and the $f_{+
         ijk'}$'s which are spin-connection (unimodular and non-unimodular) fluxes,
         \item[-] their  ``non-geometric'' counterparts: the $f_{+ ij'k'}$'s which are the
         T-dual NS--NS fluxes and the  $f_{+ i'j'k'}$'s which are
         the T-dual spin-connection fluxes;
\end{description}
    \item $f_{- IJK},\xi_{- L}$: 232 magnetic-dual parameters
    which include similarly
       \begin{description}
         \item[-] the NS--NS- and   spin-connection-like fluxes $f_{- ijk}$ and $f_{-
         ijk'}$,
         \item[-] their   T-duals $f_{- ij'k'}$ and $f_{-
         i'j'k'}$.
\end{description}
    \end{itemize}
We know that the  parameters $f_{+ ijk}$, $f_{+ ijk'}$ and the
corresponding $\xi_{+i}$ have a clear higher-dimensional geometric
interpretation as ten- or eleven-dimensional supergravity
reductions. The other parameters can also be switched on though,
leading to a well-defined gauged supergravity: the number of
degrees of freedom is not altered but the algebra, its
$SL(2,\mathbb{R}) \times SO(6,n)$ embedding, the charges and the
potential are. However, it is not clear that a higher-dimensional
set up exists, which could reproduce all these gaugings upon
dimensional reduction. This set up might simply not exist, or be a
purely string-theory non-geometric vacuum, or some more exotic
construction sitting between supergravity and string theory, like
a double-torus compactification\cite{Hull:2004in,Shelton:2005cf,
Dabholkar:2005ve,Hull:2006va,Hull:2007jy,Dall'Agata:2007sr}.
Examples exist where this is indeed suspected to happen. These
include four-dimensional $N=4$ gauged supergravities that admit de
Sitter vacua \cite{deRoo:2002jf} and are build using de
Roo--Wagemans phases. As we pointed out in Sec. \ref{sec:gausugr},
de Roo--Wagemans phases are equivalent to switching on both
electric ($f_{+ IJK}$) and magnetic ($f_{- IJK}$) gauging
parameters. Since it seems hard to oxidize de Sitter vacua to
higher-dimensional supergravity \cite{Maldacena:2000mw}, we might
conclude that the higher-dimensional origin of gauged
supergravities with both electric and magnetic parameters is hard
to achieve, without excluding its realization at the string level.

\section*{Acknowledgements}

The author would like to thank Jean-Pierre Derendinger and
Nikolaos Prezas for nice collaboration, and Thomas Van Riet for
stimulating scientific discussions during the Valencia meeting.
This research was partially supported by the EU under contracts
MEXT-CT-2003-509661, MRTN-CT-2004-005104 and MRTN-CT-2004-503369,
by the French Agence Nationale pour la Recherche and by the Swiss
National Science Foundation.

\end{document}